\documentclass[prl, aps, twocolumn, superscriptaddress, nofootinbib, tightenlines, nobibnotes]{revtex4-1}

\usepackage{bigints}
\usepackage{amssymb}
\usepackage{graphicx}
\usepackage{dcolumn}
\usepackage{bm}
\usepackage{multirow}
\usepackage{hyperref}
\usepackage{comment}
\usepackage{xcolor}
\usepackage{float}

\usepackage{soul}

\hypersetup{
colorlinks=true,
linkcolor=blue,
linktoc=page,
citecolor=blue,
urlcolor=blue
}

\everymath{\displaystyle}

\usepackage{tikz}
\usetikzlibrary{patterns,arrows,shapes,arrows,intersections,decorations}

\begin{document}

\preprint{APS/123-QED}

\title{Topological waves in the continuum in magnetized graphene devices}

\author{Alexandru Ciobanu}
\email{alexandru.ciobanu@tecnico.ulisboa.pt} 
\affiliation{GoLP $\vert$ Instituto de Plasmas e Fus\~ao Nuclear, Lisboa, Portugal}
\affiliation{Instituto Superior T\'ecnico, Lisboa, Portugal}

\author{Pedro Cosme}
\affiliation{GoLP $\vert$ Instituto de Plasmas e Fus\~ao Nuclear, Lisboa, Portugal}
\affiliation{Instituto Superior T\'ecnico, Lisboa, Portugal}

\author{Hugo Ter\c{c}as}
\email{hugo.tercas@tecnico.ulisboa.pt} 
\affiliation{GoLP $\vert$ Instituto de Plasmas e Fus\~ao Nuclear, Lisboa, Portugal}
\affiliation{Instituto Superior T\'ecnico, Lisboa, Portugal}

\begin{abstract}
We show that topological waves at the interface between two magnetic domains in a graphene device are possible. First, we consider the case of a linear relation between the applied gate voltage and local density in the channel and, secondly, we investigate the effect of non-local Coulomb interactions. We obtain two distinct edge modes for each interaction type: a Yanai mode with rotational flow and dispersion relation that extends to infinite wave-number, and a Kelvin mode with purely longitudinal flow and bound dispersion relation. The scattering matrix concept is applied to verify the infinite frequency regime of the spectrum, and the bulk-edge correspondence principle is satisfied if one takes into account the Kelvin modes that merge with an imaginary cut of the bulk band.

\end{abstract}

 %Use showkeys class optio
\maketitle

%%%%%%%%%%%%%%%%%%%%%%%introdução%%%%%%%%%%%%%%%%%%%%%%%%%%%%%

{\it Introduction.}---The process of understanding protected edge modes from topological considerations has reached a tremendous success in different physical platforms, including electronic condensed matter systems \cite{PhysRevLett.49.405,PhysRevB.23.5632,PhysRevB.25.2185,RevModPhys.82.3045}, mechanical oscillators \cite{PhysRevLett.115.104302}, photonics \cite{PhysRevA.78.033834,PhysRevX.5.031011,PhysRevX.9.011037}, acoustics \cite{PhysRevLett.103.248101,PhysRevLett.123.054301}, active matter \cite{PhysRevX.7.031039}, molecular spectra \cite{PhysRevLett.85.960} and superfluids \cite{super}. Systems with a broken time-reversal symmetry are described in terms of a topological invariant, the so-called Chern number \cite{haldane_2008, raghu_2008, Wang_2009, lu_2014, Benalcazar_2017, Khanikaev_2017, ozawa_2019}. For fermionic systems, the latter determines the quantized Hall conductivity \cite{laughlin_1981, thouless_1982, halperin_1982, haldane_1988}. A second generation of experiments reporting on the observation of Hall effect in graphene \cite{Zhang_2005, Novoselov_2007}, topological insulators \cite{Bhardwaj_2021} and antiferromagnetic films \cite{Zhang_2019} has renewed interest around the topological classification of gapped systems. The investigation of topological invariants culminated in the celebrated bulk-edge correspondence \cite{hatsugai_1993a, hatsugai_1993b, Hasan_2010}, stating that if a Chern number exists for gapped system (bulk), then topologically protected modes appear at the boundaries (edge), and vice-versa. In crystalline systems, covering a extensive class of problems in condensed matter, such modes are confined at the boundary, robust against defect scattering and coincide in number with the bulk Chern number.  \par
Continuous systems with time-reversal broken symmetry, such as geophysical and astrophysical  flows \cite{Wang_2015, Perrot_2018, Perrot_2019}, shallow waters \cite{Monteiro_2021, Graf_2021}, polar active fluids \cite{Shankar_2017, Souslov_2017}, and plasmas \cite{Gao_2016, Jin_2016, Jin_2019, Fu_2021, Silveirinha_2022} are also known to support topologically protected edge modes. However, continuous media do not support a compact torus (Brillouin zone) in which the quasi-momenta live. As a result, the classification of the bulk in term of a Chern number becomes critical, since the latter may diverge. One must adopt compactification strategies allowing for the regularization of momenta. For example, Silveirinha suggested that the introduction of a cut-off regularizes the Chern number, allowing for the application of the bulk-edge correspondence \cite{Silveirinha_2015, Silveirinha_2016}, and has recently established a link between topology and fluctuation-electrodynamics \cite{PhysRevX.9.011037}. In Refs. \cite{PhysRevLett.122.128001}, Souslov et al. show that odd viscosity naturally compactifies the bulks of the acoustic waves, without the need of introducing cut-offs. Also, in the work of Tauber et al. \cite{tauber_delplace_venaille_2019}, a solution to the bulk-edge correspondence in terms of ghost modes has been proposed; in a subsequent work, the authors make use of a scattering matrix argument \cite{PhysRevResearch.2.013147}.
\begin{figure}[t!]
    \centering
    \includegraphics[width = 0.8\linewidth]{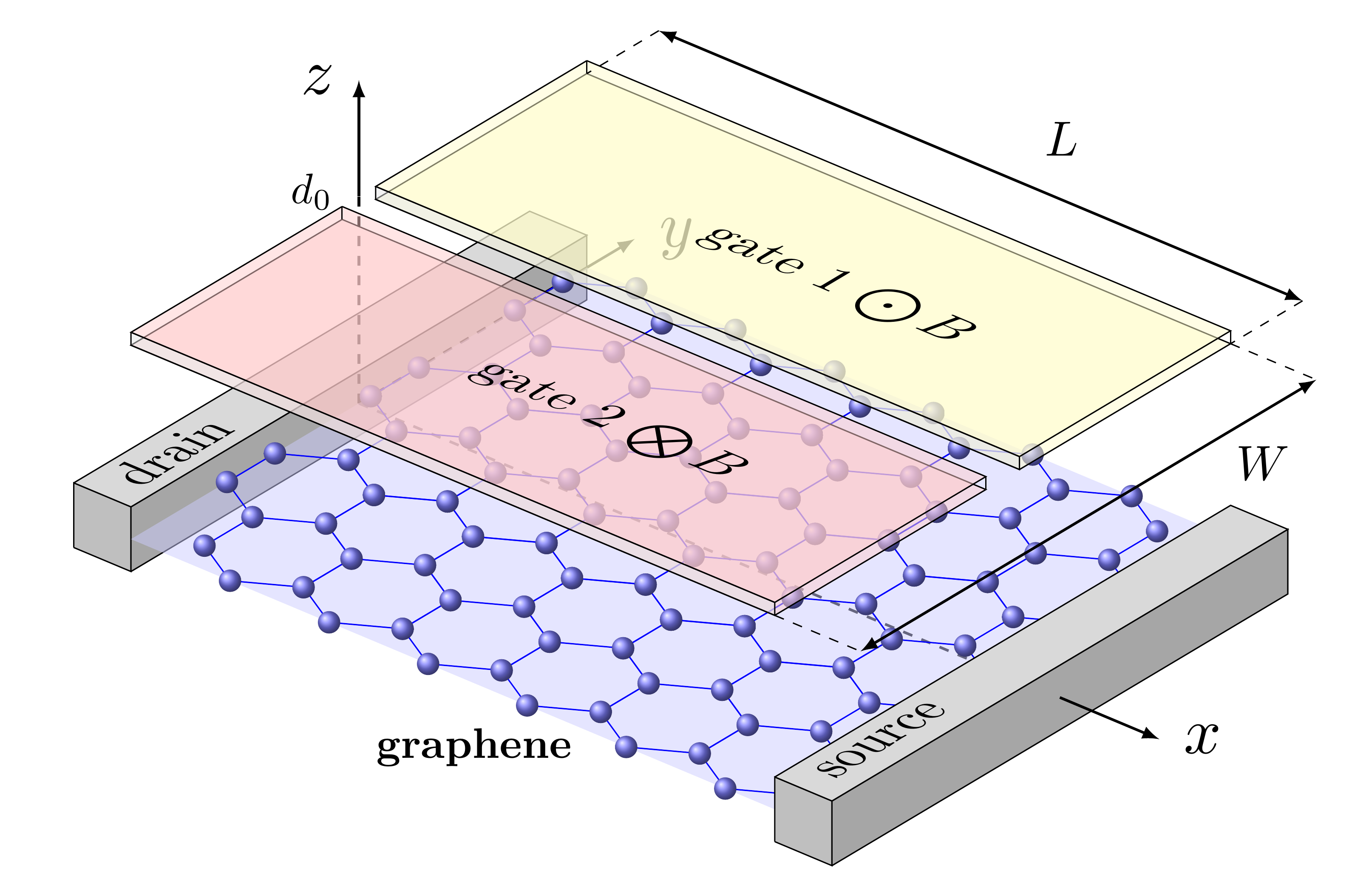}
    \caption{(color online) Schematics of the dual-gate graphene field-effect transistor. A graphene sheet, with dimensions $L \times W$, is located in the $z = 0$ plane. The magnetized dual-gate plate produces two different magnetic domains, characterizing by Chern number $C=\pm 2$ in the presence of odd viscosity.}
    \label{fig:setup}
\end{figure}
\par
In this Letter, we investigate topological modes in two-dimensional graphene plasmas. We show that topological waves emerge at the interface of two magnetic domains of a graphene device. Due to the electrostatic nature of the modes, the boundary conditions at the interface are determined by the continuity of the longitudinal component of the electric field, in contrast to the boundaries usually imposed in chiral active fluids, for example \cite{Souslov_2021}. Two different cases are studied: first, by making use of a graphene field-effect transistor (FET)
configuration, we control the electrostatic potential with the help of a gate (screened case), as illustrated in Fig. (\ref{fig:setup}), resulting in a local relation between the potential and the electron density, $\phi({\bf r})\sim n({\bf r})$. In a second configuration (unscreened case)-- which has not been addressed so far to the best of our knowledge -- we remove the electrostatic and let the electrons respond to their self-consistent Coulomb interaction, which results in a non-local potential $\phi({\bf r})\sim \int V({\bf r}-{\bf r}')n({\bf r}') d{\bf r}'$. In both scenarios, we reveal the emergence of topological waves of the Kelvin ($\omega\sim q$) and Yanai ($\omega\not\sim q$) types \cite{Delplace_2017}, both being chiral edge modes: the former bridging the two bulk by crossing the gap; while the latter connects the two bulk bands at infinity without crossing the gap. 

%Hydrodynamic simulations are performed with \textit{TETHYS} software \cite{doi:10.1021/acsphotonics.0c00313} to excite the calculated velocity and density profiles. Lastly, the scattering matrix concept is applied to the case of two sub-systems in order to verify the merging of Yanai modes with the bulk bands and the existence of ghost modes at infinite frequencies.

{\it Graphene hydrodynamical model.}--- We start by presenting the equations governing the dynamics of electrons in magnetized graphene in the hydrodynamical regime \cite{doi:10.1021/acsphotonics.0c00313} 
\begin{equation}
\begin{aligned}
 \frac{\partial n}{\partial t} =& -\bm \nabla \cdot(n\bold{v}) ,\\
 \frac{\mathcal{D}}{\mathcal{D} t}\bold{v} =& -\frac{\bm \nabla P}{nm_e} - \frac{e}{m_e} \left(-\bm\nabla \phi + \bold{v}\times \bold{B}\right) + \nu^o\bold{\nabla}^2\bold{v^*}, 
\end{aligned}
\label{eom}
\end{equation}
where $\phi$ is the electrostatic potential (to be specified below), $\bold{v}^* = \left(v_y,-v_x\right)$, $P=\hbar v_F\sqrt{\pi n^3}/3$ is the Fermi pressure (in the limit of high chemical potential assumed in this work). The operator $\mathcal{D}/\mathcal{D}t \equiv \partial/\partial t +\left(\bold{v}\cdot\bm\nabla\right)-(1/2)\bm\nabla\cdot\bold{v}$ containing the anomalous convective term, and the effective hydrodynamical mass $m_e=\hbar \sqrt{\pi n}/v_F$ are a peculiarity of the Dirac (massless) dispersion of the electrons \citep{Cosme_2020}. The odd-viscosity reads \cite{PhysRevLett.122.128001} 
\begin{equation}
\nu^o = \frac{k_BT}{2m_e\omega_B}\left( \frac{4\omega_B^2\tau^2}{1+4\omega_B^2\tau^2}\right),
\end{equation}
with $\tau$ being a characteristic relaxation time and $\omega_B =-eB/m_e$ the cyclotron frequency. Note that $\{\omega_B, \nu^o \} \rightarrow  -\{\omega_B, \nu^o \}$ when inverting the direction of the $B$.
\par
%Dwd
\begin{figure}[t!]
    \centering
    \includegraphics[width = 0.8\linewidth]{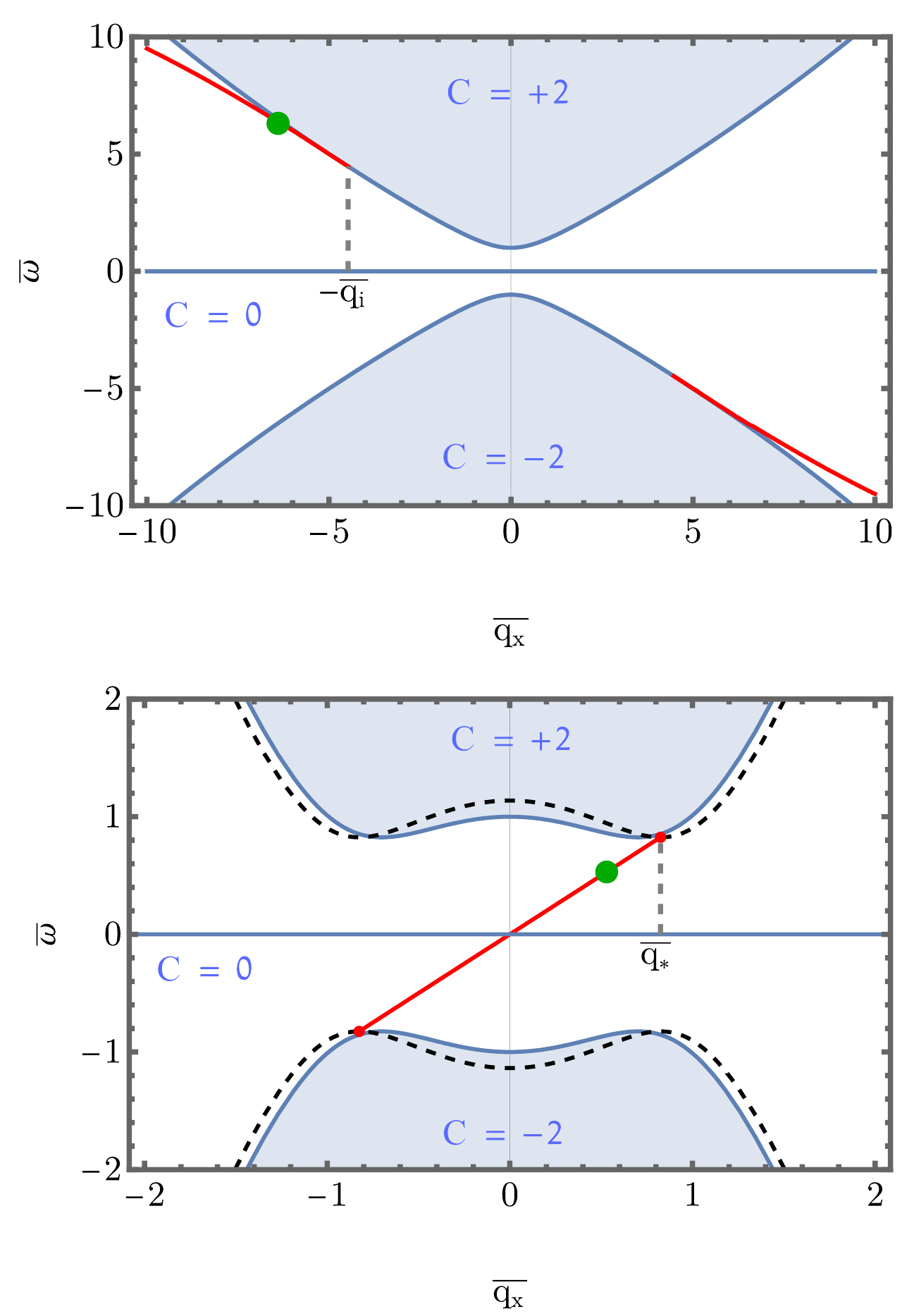}
    \caption{(color online) Dispersion relation for the screened graphene plasma with two magnetic bulk modes (shadowed regions) and topological modes (red lines). Top panel: Yanai mode dispersion relation for $m = 0.05$. Bottom panel: Kelvin mode dispersion for $m = 1.15$. The imaginary cuts ${\omega}_\pm(\pm {q}_x, i/2m)$ is given by the dashed line, showing the point where the Kelvin modes terminate ($q_x= q_*$). The green dots are the picked excitation frequencies in the simulations of Fig. (\ref{fig:sim_screened}).}
    \label{fig:screened_dispersion}
\end{figure}
\textit{Topological waves in screened graphene.}--- Within the gradual channel approximation, the potential $\phi$ in the plasma is approximated by \cite{1444418}
%dwdw
\begin{equation}
    \phi(\bold{r}) =  -\frac{ed}{\epsilon}n(\bold{r}), \quad {\rm with} \quad  \bold{r}=(x,y),
\end{equation}
%dwd
where $d$ is the gate-channel distance and $\epsilon$ the permittivity of the substrate. The linearization of \eqref{eom} for vector states of the form  $\psi\equiv (n/n_0,v_x/c, v_y/c)\sim e^{i(\omega t - q_xx - q_yy)}$, with $n_0$ the equilibrium density and $ c =\sqrt{v_F^2 + (n_0e^2d)/(m_e\epsilon) }$ denoting the plasmon sound speed, yields the eigenvalue problem $\mathcal{H}\psi =\omega\psi$, where the Hamiltonian is defined as
%dw
\begin{equation}
\mathcal{H}=\begin{pmatrix} 0 &{q}_x &{q}_y\\ {q}_x &0 & -i(1 - m{q}^2) \\{q}_y& i(1 - m{q}^2) & 0
\end{pmatrix} 
\end{equation}
%wd
where we made use of the adimensionalization $qc/\omega_B\to q $, $\omega_B\nu^o/c^2 \to m$ and $\omega/\omega_B\to \omega$ that we keep until otherwise noticed. This yields the spectrum ${\omega}_\pm= \pm \sqrt{{q}^2 + (1-m{q}^2)^2}$, and the corresponding eigenstates
%dw
\begin{equation}
\psi_\pm({q}_x,{q}_y) = \frac{\lambda_\pm}{\sqrt{2}|{q}|}\begin{pmatrix}
{q}^2/{\omega}_+\\ {q}_x \mp i{q}_y(1 -m{q}^2)/{\omega}_+ \\ {q}_y \pm i{q}_x(1 -m{q}^2)/{\omega}_+  \end{pmatrix}\label{state},
\end{equation}
%dwd
with $\lambda_\pm = |{q}|^{-1}({q}_x \pm i{q}_y)$. The Chern number is defined as the flux of the \textit{Berry curvature} \cite{Berry_1984, Souslov_2017, ozawa_2019}
%dwd
\begin{equation}
    C_\pm = \int (\nabla_q \times \bold{A_\pm}) d^2\bold{q},\label{chern_form}
\end{equation}
%dw
where $\bold{A}_\pm = -i\psi_\pm^\dagger.\nabla_q \psi_\pm$ is the \textit{Berry connection}.
\begin{figure}[t!]
    \centering
    \includegraphics[width = 1.0\linewidth]{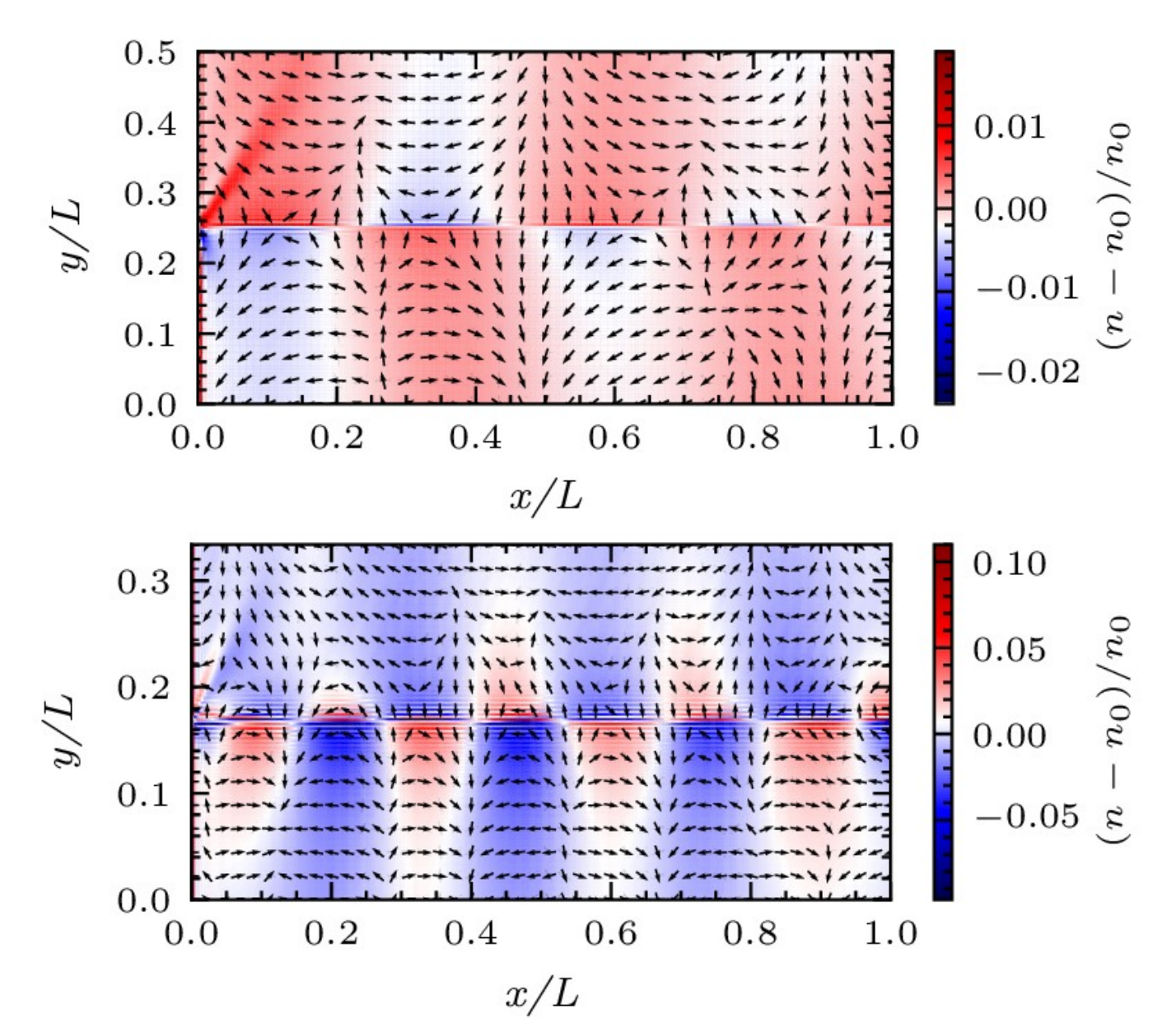}
    \caption{(color online) Numerical simulations of the topological modes for screened a graphene plasma. Top:  Yanai dispersion relation, obtained for $m = 0.1$, and excited at the frequency $\omega=6.5 \omega_B$. Bottom: Kelvin mode dispersion obtained for $m = 0.95$ and excited at $\omega=0.65\omega_B$. In both panels, the arrows display the velocity field $(v_x,v_y)$, while the false color map depicts the density contrast $(n-n_0)/n_0$.}
    \label{fig:sim_screened}
\end{figure}
From \eqref{state}, one finds $C_\pm = \pm (1 + \text{sign}(m)) = \pm 2.\label{chern_spec}$ for the upper and bottom bulks, respectively (see Fig.\ref{fig:screened_dispersion}). In order to find edge modes at the interface of the two magnetic domains, we consider perturbations of the form $\psi = \psi(y)e^{i(\omega t - q_xx)}$ in \eqref{eom}, 
\begin{equation}
\begin{array}{c}
  \left(m (\partial_{{y}}^2 - {q}_x^2) -\frac{{q}_x}{{\omega}}\partial_{{y}} + 1 \right)v_y   = -\frac{i}{{\omega}} ({\omega}^2 - {q}_x^2)v_x,\\\\
      \left(m(\partial_{{y}}^2 -{q}_x^2 ) + \frac{{q}_x}{{\omega}}\partial_{{y}} + 1\right)v_x = \frac{i}{{\omega}}({\omega}^2 +\partial_{{y}}^2)v_y 
\end{array}
    \label{init1}.
\end{equation}
We proceed to solve \eqref{init1} in each magnetic domain and apply fluid-fluid boundary conditions \cite{PhysRevE.104.014603}
%Dwd
\begin{equation}
    \bold{v}(0^+)= \bold{v}(0^-), \quad \partial_y\bold{v}(0^+) = \partial_y\bold{v}(0^-).
    \label{bound_velocity}
\end{equation}
%dwd
We obtain two edge modes, as depicted in Fig.\ref{fig:screened_dispersion}: a Kelvin mode (bottom panel), a purely longitudinal wave ($v_y = 0$) that crosses the spectral gap. It consists of a dispersionless wave $ \omega = q_x$ and fades away into the bulk as $(v_x(y), n(y))\sim e^{-\vert y\vert/2m}\sin( q_o  y)$, where ${q}_o =\sqrt{|1-4m+ 4m^2{q_x}^2|}/2m$.  This mode terminates at $ q_*= \sqrt{m^{-1} -(4m^2)^{-1}}$ (see dashed line in Fig.\ref{fig:screened_dispersion}), still before hitting the bulk at $ q_i=\sqrt{1/m}$. Secondly, we find a Yanai mode (top panel), associated to a rotational flow ($v_y\sim \pm i v_x$), which is a evanescent, dispersive wave ($ \omega\neq  q_x$), defined in the region $ q_i <\vert  q_x \vert < \infty $, and fades into the bulk as $(v_x(y), n(y))\sim A_1 e^{-s_1\vert y\vert}+A_2 e^{-s_2\vert y\vert} $, with $s_{1,2}$ are functions of $ q_x$. At infinity, $ q_x \to \pm \infty$, the Yanai frequency scales as $ \omega \to \pm m^{-1}$. To corroborate our analytical findings, we proceed to a numerical solution of \eqref{eom} with the help of TETHYS \cite{Cosme_2023, Cosme_2021} $-$ a hydrodynamical simulator for graphene plasmons $-$ by exciting the system with a selected frequency, represented by the green dots in Fig.\ref{fig:screened_dispersion}. The numerical simulations reporting the main features of Kelvin and Yanai modes shown in Fig.\ref{fig:sim_screened}.

%%%%%%%%%%%%%%%%%%%%%%%%%%%%%%%%%%%%%%%%%%%%%%%%%%%%%%%%%%%%%%%%%%%%%%%%%%%%%%%%%%%%%%%%%%%%%%%%%%%%%%%%%%%%%%%%%%%%%%%%%%%%%%%%%%%%%%%%%%%%%%%%%%%%%%%%%%%

\par
\begin{figure}[t!]
    \centering
    \includegraphics[width = 0.8\linewidth]{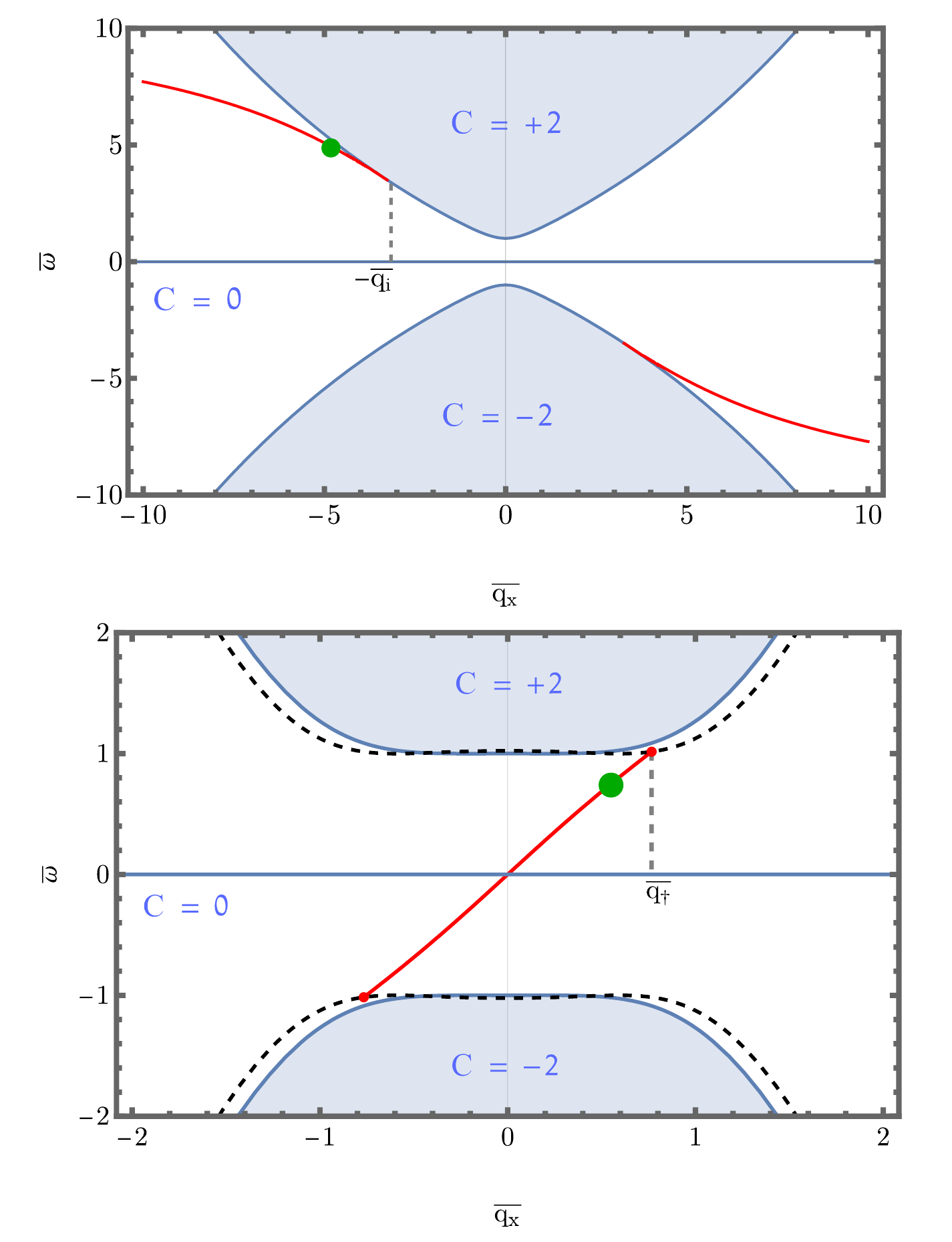}    
    \caption{(color online) Dispersion relation for the unscreened graphene plasma with two magnetic bulk modes (shadowed regions) and topological modes (red lines). Top panel: Yanai mode dispersion relation for $m = 0.05$. Bottom panel: Kelvin mode dispersion for $m = 1.15$. The imaginary cuts ${\omega}_\pm(\pm{q}_x, i/2m)$ is given by the dashed line, showing the point where the Kelvin modes terminate ($q_x= q_*$). The green dots are the picked excitation frequencies in the simulations of Fig. \ref{fig:sim_unscreened}.}
    \label{fig:unsscreened_dispersion}
\end{figure}
\begin{figure}[ht!]
    \centering
    \includegraphics[width = 1.0\linewidth]{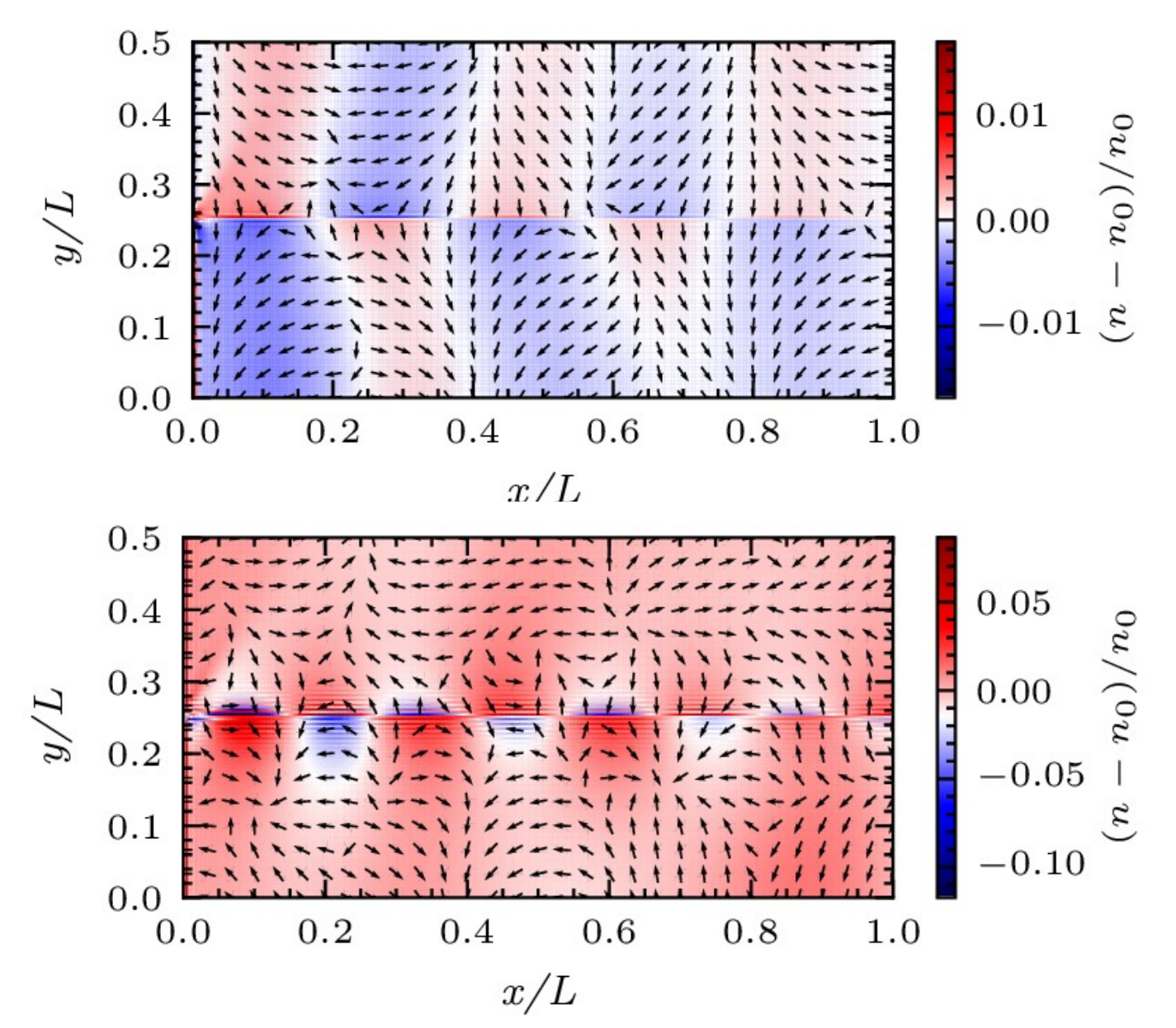}
    \caption{(color online) Numerical simulations of the topological modes for screened a graphene plasma. Top:  Yanai dispersion relation, obtained for $m = 0.1$, and excited at the frequency $\omega=5.0 \omega_B$. Bottom: Kelvin mode dispersion obtained for $m = 0.95$ and excited at $\omega=0.85\omega_B$. In both panels, the arrows display the velocity field $(v_x,v_y)$, while the false color map depicts the density contrast $(n-n_0)/n_0$.}
    \label{fig:sim_unscreened}
\end{figure}

\textit{Topological waves in unscreened graphene.}--- We now investigate the case of non-local interactions, so that $\phi$ satisfies the Poisson equation
%dw
\begin{equation}
(\nabla^2-\lambda^{-2}) \phi(x,y,z) = \frac{e}{\epsilon}n(x,y)\delta(z),
 \label{ungated}
\end{equation}
%ewe
where a length $\lambda$ is added to assure the infrared compactification of the topology, $\bold{q} \to 0$. In the Fourier space, the latter yields
\begin{equation}
    \phi(q) = -\frac{ e}{2\epsilon}\frac{n(q)}{\sqrt{q^2+\lambda^{-2}}}\equiv -\frac{ e}{2\epsilon}n(q)\xi,
\end{equation}
Eventually, we set $\lambda \to \infty$ (i.e. $\xi\to q^{-1}$) in the calculations whenever possible. Since the plasma dispersion for ungated electrons in 2D is no longer linear we now change our choice of normalization to the Fermi velocity $v_F$ instead, i.e. 
where the wavevector is normalized $v_Fq/\omega_B\to q$. The Hamiltonian obtained from linearization of \eqref{eom} and \eqref{ungated} together now reads 
\begin{equation}
    \mathcal{H}=\begin{pmatrix} 0 &{q}_x &{q}_y\\ \left(1 + {g}_e\xi\right){q}_x &0 & -i(1- m{q}^2) \\\left(1 + {g}_e\xi\right){q}_y& i(1-m{q}^2) & 0
    \end{pmatrix}, 
\end{equation}
%dwd
where we have defined $g_e=e^2n_0/2m_e\epsilon$, and the corresponding spectrum now yields ${\omega}_\pm= \pm \sqrt{ {q}^2(1+{g}_e{\xi}) + (1- m{q}^2)^2 }$. From the corresponding eigenstates $\psi_\pm$, we also obtain the Chern numbers as $C_\pm = \pm (1+ \text{sign}(m))$. Assuming $\phi = \phi(y)e^{i(\omega t-q_xx)}$ in \eqref{ungated} and substituting into \eqref{eom}, we obtain 
%dwd
\begin{widetext}
\begin{equation} 
  \begin{aligned}
     \left(m (\partial_{{y}}^2 - {q}_x^2) -\frac{{q}_x}{{\omega}}\partial_{{y}} + 1 \right)v_y &   -i {q}_x\frac{{g}_e}{\pi{\omega}} \int d\zeta \mathcal{V}(\zeta)K_0\left(\frac{|{y}-\zeta|}{\xi}\right) = -\frac{i}{{\omega}} ({\omega}^2 - {q}_x^2)v_x, \\
     \left(m (\partial_{{y}}^2 - {q}_x^2) +\frac{{q}_x}{{\omega}}\partial_{{y}} + 1 \right)v_y &  +  \frac{{g}_e}{\pi{\omega}}\int d\zeta\mathcal{V}(\zeta)\partial_{{y}}K_0\left(\frac{|{y}-\zeta|}{\xi}\right) =\frac{i}{{\omega}} ({\omega}^2 +\partial_{ y}^2 )v_y
\end{aligned}
\label{eq:4}
\end{equation}
\end{widetext}
%dwd
where $\mathcal{V}(\zeta) \equiv {q}_xv_x(\zeta) +i (\partial_{{y}}v_y)\vert_{y=\zeta}$ and $K_0(.)$ is the modified Bessel function of the second kind. In order to find an analytical decay length we also take the zeroth order approximation $\sqrt{{\lambda}^{-2} +{q}_x^2 + {q}_y^2}\rightarrow  \sqrt{{\lambda}^{-2} +{q_x}^2 }$, of the Fourier-transform of the Bessel function $\tilde{K}_0({q})$. For $m  < 1/4$, there exists a Yanai mode with dispersion relation shown in Fig.\ref{fig:unsscreened_dispersion} (top panel) that merges with the bulk band at ${q}_x = -{q}_i$, and with the typical rotational flow in the upper ($u$) and lower ($d$) domains as $v^{u,d}_x \propto \pm iv^{u,d}_y$. For $ m  > (1 + {g}_e{\lambda}/v_F^2 )/4$, there exists a Kelvin mode
with dispersion shown in Fig.\ref{fig:unsscreened_dispersion} (bottom panel),
 ${\omega} = {q}_x\sqrt{1+{g}_e\xi},\label{kelvin_ung_disp}$. The latter fades away into the bulk as an evanescent mode of the form $v_x^{u,d}(y) \propto e^{\mp{y}\sqrt{1+ {g}_e{p}}/2m}\sin\left({q}_o{y} \right)$, with $q_{o}=\sqrt{\left|1+ {g_e}\xi + 4m(m{q_x}^2-1)\right|}/2m$, valid for ${q}_x <{q}_\dagger$ implicitly defined as ${q}_\dagger = \sqrt{m^{-1} - (1+{g}_e\xi({q}_\dagger))/(4m^2)}$. The simulated velocity and density profiles are shown in Fig.\ref{fig:sim_unscreened}.
%dwd

\begin{figure}[t!]
\includegraphics[width = 0.9\linewidth]{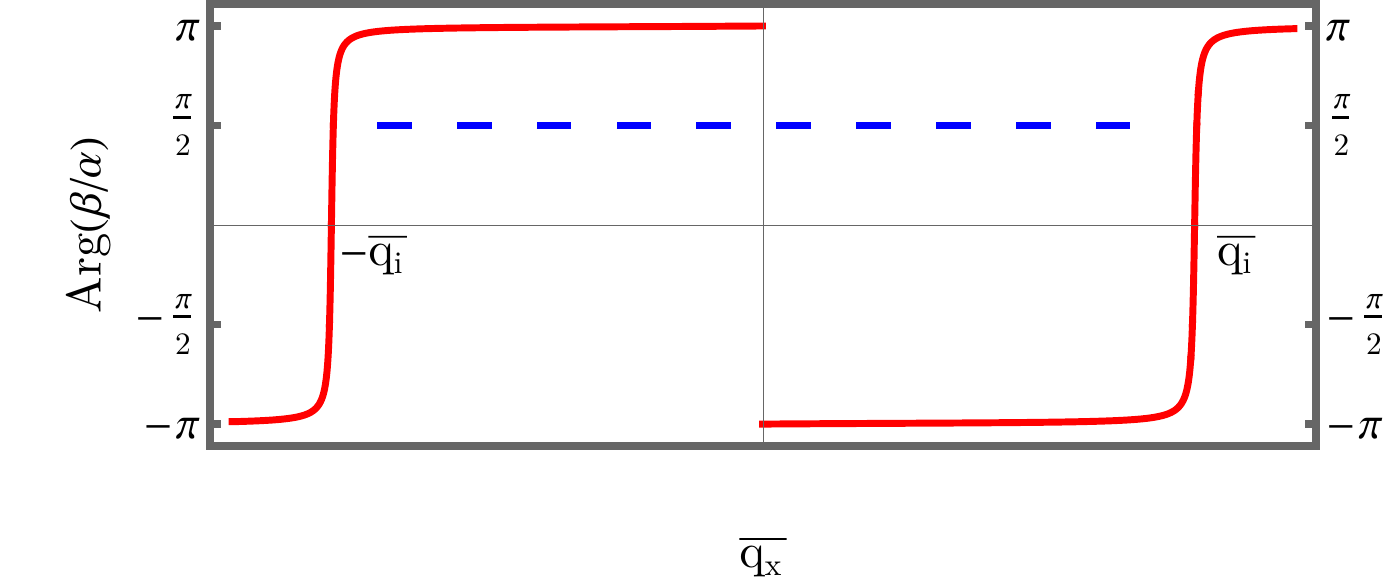}
\caption{(color online) Scattering phases from the scattering matrix theory. Scattering argument from $\psi_0$ in \eqref{state} at the bottom of the band ${\omega}_+$ for ${k}\rightarrow 0$ (solid line), where sharp jump occurs at the point the edge mode merges with the bulk band ${q}_i$, as in
Fig. \ref{fig:screened_dispersion}--top panel. Scattering argument at the top of the bulk  band, calculated with $\psi_\infty$ for ${k} \rightarrow \infty$, where no overall $2\pi$ change is present (dashed line).}
\label{fig:scattering}
\end{figure}

%dwd
\textit{Scattering matrix analysis.}--- In order to further identify the modes involved and investigate the occurrence of ghost modes, we resort to the scattering matrix formalism. % 
A scattering state $\psi_{\rm scat.}$ is constructed on each side of the interface (up, $u$, and down, $d$), from eigenstates $\psi_+$ of the bulk band with positive eigenvalue ${\omega}_+$ and evanescent solutions $\omega({q}_x,{k}) = \omega({q}_x,{q}_b)$ (here, ${q}_{b}$ is the bound-state decay rate), as \cite{PhysRevResearch.2.013147}
%dwdw
\begin{equation} 
\begin{aligned}
       \psi_{\rm scat}^{u} &=\alpha\psi^{u}(-{k})e^{i{k}{y}} + \beta\psi^{u}({k})e^{-i{k}{y}} + \gamma\psi^{u}(-{q}_b)e^{i{q}_{b}{y}}, \\
      \psi_{\rm scat}^{d} &= a\psi^{d}({k})e^{-i{k}{y}} +  b\psi^{d}(-{k})e^{i{k}{y}} + c\psi^{d}({q}_b)e^{-i{q}_b{y}},
\end{aligned}
      \label{scatt_state}
\end{equation}
%wew
where ${\omega}_\pm \rightarrow - {\omega}_\pm$ when $u \rightarrow d$.
Applying the boundary conditions in (\ref{bound_velocity}), we can get the following relation for the amplitudes
\begin{equation}
     \begin{pmatrix}\beta\\ b\end{pmatrix} =   \begin{pmatrix}
    S_{11} & S_{12}\\ S_{21} & S_{22}\end{pmatrix} \begin{pmatrix}\alpha\\ a \end{pmatrix},\label{scattering_matrix}
\end{equation}
%dwd
for some matrix elements $S_{ij}(q_x,k)$. We are interested in the amplitudes $\beta/\alpha$ for $y>0$ and $b/a$ for $y<0$ as those do not vanish in the $|y| \rightarrow \infty$ limit. Here, to fix the remaining ratios we impose $\beta/\alpha=b/a$. Thus, we obtain a general relation between amplitudes of states with Chern number $C_{u} = +2$ scattering from the top and $C_d = -2$ states scattering from the bottom. The results of the winding of the scattering matrix argument are shown in Fig.\ref{fig:scattering} for the bottom (solid red) and top (dashed blue) of the bulk band. The absence of phase jumps in $\arg(\beta/\alpha)$ indicates the absence of ghost modes. As such, the bulk-edge correspondence is achieved if a proper counting of the Kelvin and Yanai modes is in order. \par

{\it Conclusion.}--- We investigate graphene devices containing two magnetic domains in the presence of odd viscosity, and show that the latter allows for compactification, i.e. for the definition of a Chern number characterizing the bulk modes. Then, we show that the interface supports two topological modes, namely a Kelvin modes, a longitudinal wave of dispersion $\omega\sim q_x$ in the long wavelength limit and a transverse velocity profile $v_y=0$ (here, $q_x$ is the longitudinal wavevector along the interface), and a Yanai mode, a rotational, dispersive wave of dispersion $\omega\not\sim q_x$ and a velocity profile $v_y\sim \pm i v_x$. We have further shown that these modes exist for both screened and unscreened situations, governed by local and non-local electrostatic potentials respectively. To corroborate our analytical findings, we performed numerical simulations with the help of TETHYS, a hydrodynamic solver specially developed for graphene \cite{Cosme_2023, Cosme_2021, doi:10.1021/acsphotonics.0c00313}. Finally, a scattering matrix argument was employed for both magnetic domains which confirms the Yanai modes found analytically, ruling out the existence of ghost modes. As such, the bulk-edge correspondence holds if the correct number of Yanai and Kelvin modes are considered. \par
Our investigation opens the venue for a plethora of applications with topological waves in the terahertz (THz) regime, and therefore contributing for the development of topological plasmonics. In the future, the effects of nonlinearity combined with the topological protection (topological solitons, for example) deserves a close inspection, and protocols for THz sources based on the excitation of topological modes may also be in order. 

\textit{Acknowledgments.---} H.T. acknowledges Funda\c{c}\~{a}o da Ci\^{e}ncia e a Tecnologia (FCT-Portugal) through Contract No. CEECIND/00401/2018, and through the Project No. PTDC/FIS-OUT/3882/2020. P.C. acknowledges the funding provided by Funda\c{c}\~{a}o para a Ci\^{e}ncia e a Tecnologia (FCT-Portugal) through the Grant No. PD/BD/150415/2019.

\bibliography{manuscript_v5.bib}

\end{document}